# Amorphous photonic topological insulator

Peiheng Zhou[1,#], Gui-Geng Liu[2,#], Xin Ren[1], Yihao Yang[2,3,*], Haoran Xue[2,3], Lei Bi[1], Longjiang Deng[1], Yidong Chong[2,3,*], and Baile Zhang[2,3,*]

[1]National Engineering Research Center of Electromagnetic Radiation Control Materials, University of Electronic Science and Technology of China, Chengdu 610054, China.

[2]Division of Physics and Applied Physics, School of Physical and Mathematical Sciences, Nanyang Technological University, 21 Nanyang Link, Singapore 637371, Singapore.

[3]Centre for Disruptive Photonic Technologies, The Photonics Institute, Nanyang Technological University, 50 Nanyang Avenue, Singapore 639798, Singapore.

#These authors contributed equally to this work.

*Email: yang.yihao@ntu.edu.sg, yidong@ntu.edu.sg, blzhang@ntu.edu.sg

## Abstract

Photonic topological insulators (PTIs) exhibit robust photonic edge states protected by band topology, similar to electronic edge states in topological band insulators. Standard band theory does not apply to amorphous phases of matter, which are formed by non-crystalline lattices with no long-range positional order but only short-range order. Among other interesting properties, amorphous media exhibit transitions between glassy and liquid phases, accompanied by dramatic changes in short-range order. Here, we experimentally investigate amorphous variants of a Chern-number-based PTI. By tuning the disorder strength in the lattice, we demonstrate that photonic topological edge states can persist into the amorphous regime, prior to the glass-to-liquid transition. After the transition to a liquid-like lattice configuration, the signatures of topological edge states disappear. This interplay between topology and short-range order in amorphous lattices paves the way for new classes of non-crystalline topological photonic materials.

Photonic topological insulators (PTIs)[1-5] are an emerging class of photonic bandgap materials that can impart "topological protection" to photons, in the same way topological insulator materials do for electrons. The most striking feature enabled by topological protection is the existence of edge states that are protected against perturbations and defects, for which several promising applications have been identified, including robust lasers[6-8] and robust optical delay lines[9]. Topological protection originates from the topology of the underlying photonic bandstructures. The most basic class of topological insulators, Chern insulators, have integer band invariants called Chern numbers that are computed using Bloch band states, which in turn owe their existence to the discrete translational symmetry of the lattice[10-13]. Consequently, the vast majority of PTIs have been based on crystalline periodic lattices[2, 13-24] such as photonic crystals, which possess both long-range and short-range positional order. Long-range order is connected to the lattice periodicity, while short-range order is related to the regular connectivity of neighboring sites throughout the lattice[25].

Some authors have pointed out, however, that photonic topological edge states can exist even in the absence of lattice periodicity[26-28]. Of course, any local disorder can be regarded as breaking translational periodicity in the underlying lattice, and topological edge states are protected against weak disorder (for the sufficiently strong disorder, topological protection breaks down via the coupling of edge states with defect states in the bulk[26]). More surprisingly, certain lattices are topologically trivial in the absence of disorder but become "topological Anderson insulators" when the disorder is added, as recently demonstrated using a photonic lattice[28]. In discrete systems (e.g., tight-binding models), the Bott index was shown to be usable in place of the Chern number when there is no well-defined momentum space[27, 29]. The aforementioned scenarios all start from a crystalline lattice with a well-defined Brillouin zone, into which local disorder is introduced.

There are many materials in nature that exist in amorphous phases (e.g., glass, polymer, and gel) that lack any such easily-identifiable "initial" crystalline configuration. Amorphous phases of matter intrinsically lack long-range order but maintain short-range order[30]. They exhibit an interesting phenomenon known as the "glass transition", whereby an amorphous medium experiences a dramatic structural change from a glass-like phase into a liquid-like phase[31]. To date, many physical aspects of the glass transition remain poorly understood[32].

In order to study the interplay between band topology and short-range order, we have experimentally extended a Chern-number-based PTI[10-13] into the amorphous regime. Similar to previous theoretical proposals[33, 34], the amorphous PTI that we study consists of gyromagnetic rods that are arranged in computer-generated amorphous lattice patterns, and magnetically biased to break time-reversal symmetry. By performing edge/bulk transmission and near-field distribution measurements, we experimentally verify the existence of robust topological edge states in the amorphous PTIs prior to the onset of the glass transition. When the lattice undergoes the glass transition, the local site connectivity is dramatically altered, resulting in the closing of the bulk topological gap and the disappearance of the topological edge states. Although the concept of amorphous topological insulators has been theoretically proposed in condensed matter systems[35, 36], and some related features have been realized in a mechanical network of gyroscopic oscillators[37], there has never been any systematic experimental study of how band topological effects depend on short-range order (including the important role of the glass transition). This work thus enriches our understanding of topological photonic materials, and paves the way to exploring new types of photonic lattices that can host topologically protected edge states.

**Results**

Photonic lattices with different structural correlations are generated using numerical particle-packing methods previously developed in soft condensed matter physics[38-40]. The packing is conducted in a two-dimensional (2D) square unit cell with periodic boundary conditions, bidisperse discs (radius ratio 1.2 with equal distributions; see Supplementary Information). The process ends upon reaching a target packing density $\phi$ (the fraction of space covered by the discs). The configuration is then transformed into a photonic lattice by replacing the discs with gyromagnetic cylindrical rods (see Figs.1**a**,**b**; note that the crystalline lattice is built on a triangular lattice not generated by the packing method). We define a disorder index (DI) as DI = $(\phi_{max} - \phi)/\phi_{max}$, where $\phi_{max} = 0.9069$ is the densest possible packing density (corresponding to a triangular lattice). As defined, the DI is positively related to the amount of disorder in the lattice.

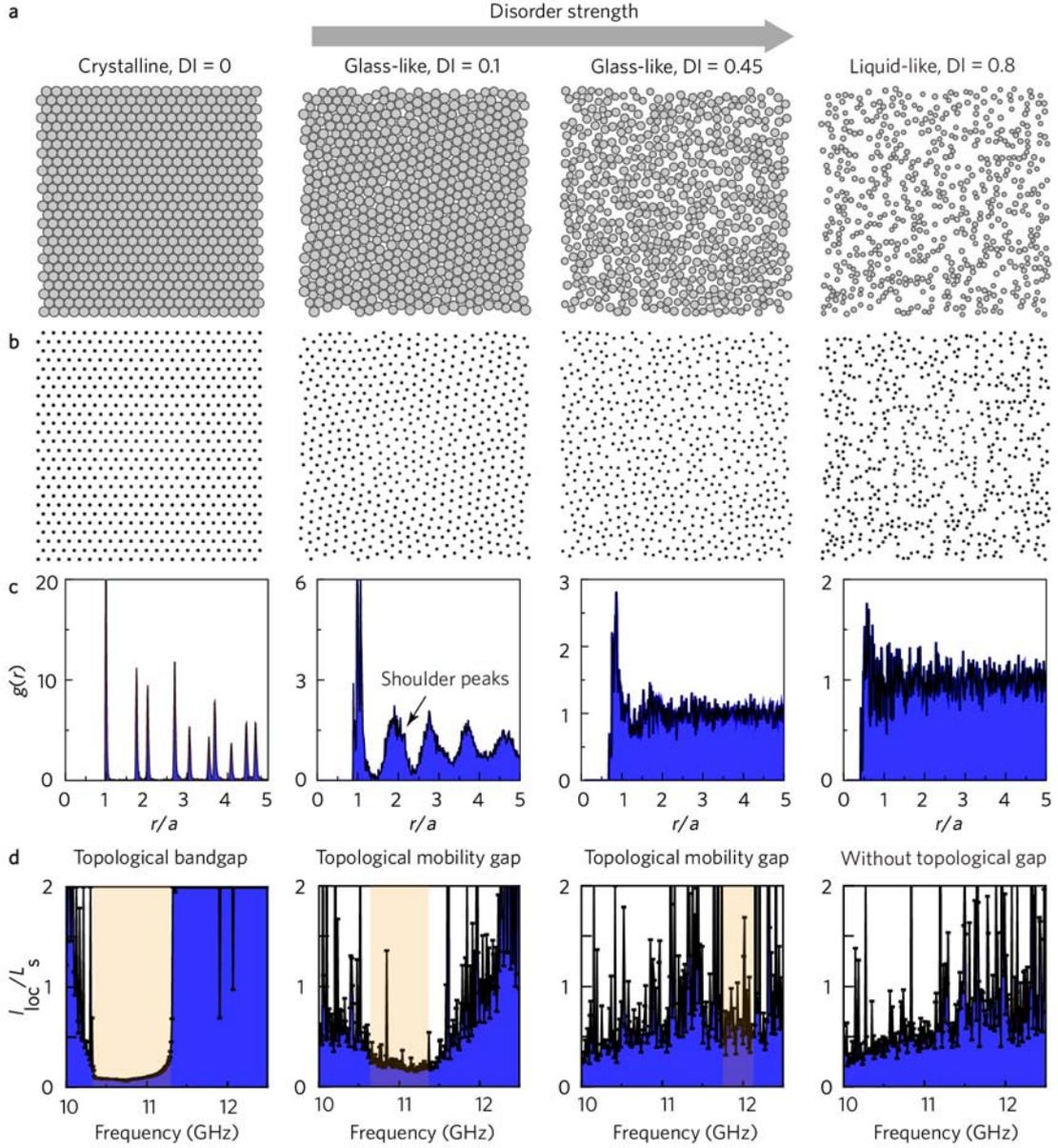

**Figure 1. Transition of photonic lattices with increasing disorder. a-b** Particle patterns (a) and the corresponding photonic lattices (b) with different structural correlations. The DI = 0 case is a triangular lattice. The glass-like lattices with strong short-range order have DI = 0.1 and 0.45. The liquid-like lattice with weak short-range order possesses DI = 0.8. **c** Pair correlation function $g(r)$ for the different lattices. **d** Numerically-calculated localization lengths for the photonic lattices. The orange regions are the frequency windows in which topological edge states can be observed. The calculation details can be found in the Supplementary Information.

Next, we examine the pair correlation function

$$g(r) = \frac{a^2}{4N\pi r^2} \sum_{i=1}^{N} \sum_{j \neq i}^{N} \langle \delta(r - r_{ij}) \rangle \tag{1}$$

where $r$ is the distance between a pair of gyromagnetic rods, $N$ is the number of rods, and

$a = L/\sqrt{N}$, where $L$ is the size of the lattice system. This quantifies the degree of structural correlation in the lattice and has been extensively employed to characterize amorphous phases[41]. The pair correlation functions for different lattice structures are plotted in Fig. 1c. For DI = 0, $g(r)$ shows sharp peaks in the whole $r$ range. For DI = 0.1, the first peaks in $g(r)$ split into subpeaks due to the bidisperse packing; the shoulders of the second peak indicate local clustering, a common phenomenon in amorphous materials[42]; the other peaks progressively damp away at $r/a > 3$, indicating the lack of long-range order. For DI = 0.45, the short-range order decreases, and the first $g(r)$ peak is less than half of the counterpart in the DI=0.1 case. For the weakly-correlated lattice with DI = 0.8, there is only one visible peak and $g(r) \sim 1$ over most of the range, indicating weak short-range order (i.e., a liquid-like lattice configuration).

These variations in lattice properties can have significant impacts on band topological phenomena. In the following content, we will start with the topological bandgap of a crystalline PTI corresponding to a DI=0 lattice (left panel in Fig. 1d). We will demonstrate that this topological gap persists for amorphous lattices up to DI= 0.45 (middle panels in Fig. 1d), before the glass transition. After the glass transition, the topological frequency gap closes (right panel in Fig. 1d).

The crystalline PTI corresponds to a triangular lattice with lattice constant $a$ = 17.5 mm, with the gyromagnetic rods having radius 2.2 mm. (See Supplementary Information for detailed system parameters.) Its bandstructure is shown in Fig. 2(a): there is a bandgap between the second and third transverse magnetic (TM) bands. From the Bloch functions, we compute the gap Chern number to be $C_p = 1$[34].

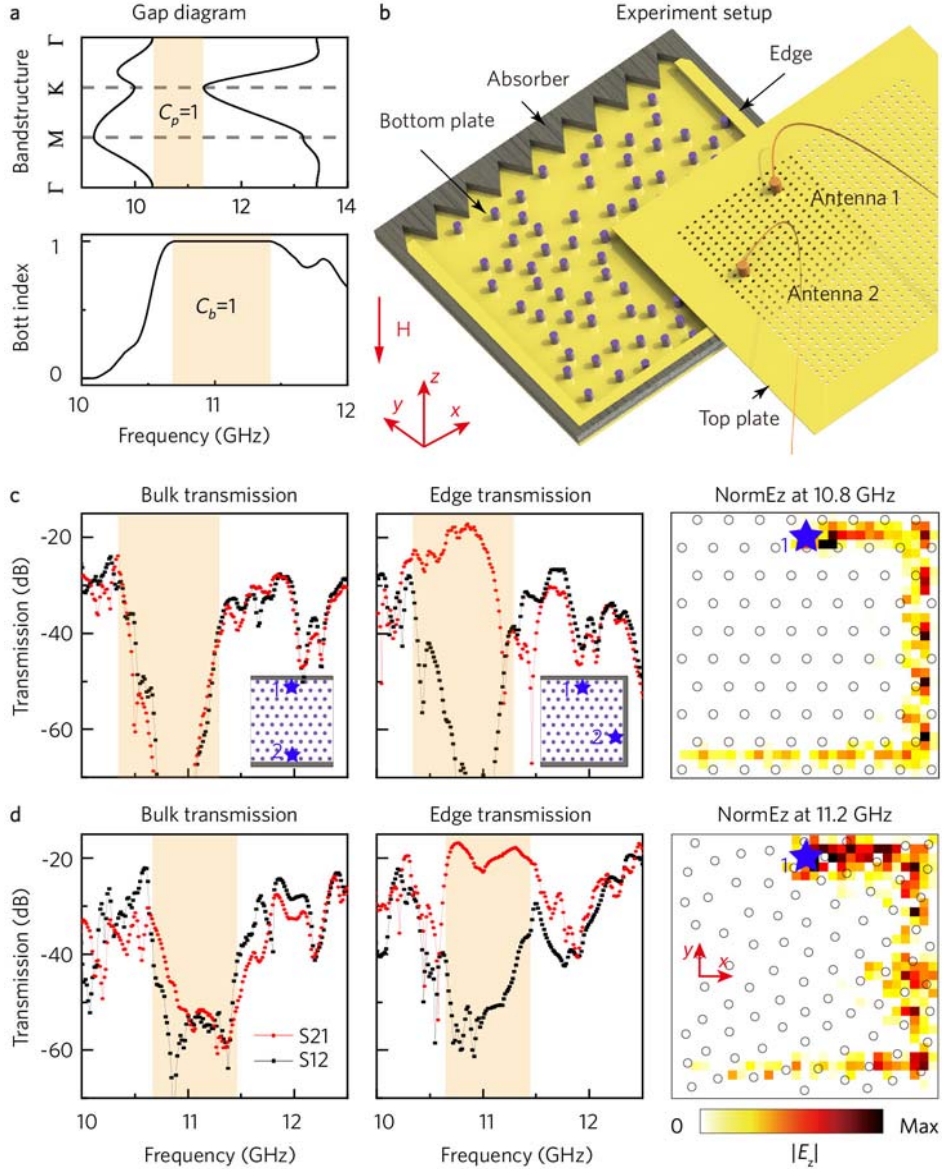

**Figure 2. Observation of topological states in an amorphous PTI. a** Numerically-calculated bandstructure of the crystalline PTI (upper panel, DI = 0) and Bott index of the amorphous lattice (lower panel, DI = 0.1). The orange region denotes the frequency window corresponding to the topological gap. **b** Schematic of the experimental setup. The top plate contains cylindrical holes of radius 1 mm. The probe and source dipole antennas (1 and 2) are inserted into the waveguide through these holes. Three sides of the waveguide are wrapped with metal walls acting as perfect electric conductor (PEC) boundaries. The other side is covered by microwave absorbers. **c** Measured S21 and S12 transmissions of bulk and edge states, and $|E_z|$ field distribution in the crystalline PTI. Insets: schematics of the experimental setup showing amorphous lattice (purple dots) and metal boundaries (grey bars) for the bulk and edge measurements. The source (1) and probe (2) antennas are indicated by blue stars. **d** Measured S21 and S12 transmissions for bulk and edge states, and $|E_z|$ field distribution in the amorphous PTI.

To characterize the present photonic lattice as well as the others, we use the experimental setup shown in Fig. 2**b**. The sample lies in a copper parallel-plate waveguide. Gyromagnetic ferrite cylindrical rods are placed on the bottom plate, and an external static magnetic field of $B = 0.2$ T is applied along the negative z direction. The sizes of all samples are tailored to be $9a \times 9a$, in order to fit our apparatus. The parallel-plate waveguide has a spacing of 4 mm, which supports only the fundamental TM mode below 37.5 GHz. To facilitate the excitation and measurement of electromagnetic fields inside the waveguide, a square array of air holes are drilled through the top plate. As the diameter of these holes is smaller than 1/15 the operational wavelength, they have a negligible influence on the electromagnetic modes in the waveguide, as verified via first-principles calculations (see Supplementary Information).

To study the bulk states, source and probe antennas are placed near the top and bottom edges (near the metallic walls), while the left and right sides of the photonic lattice are wrapped with microwave absorbers (Fig. 2**c**). The measured forward (S21) and backward (S12) transmission through the bulk shows a dip from 10.4 GHz to 11.3 GHz, indicating a bandgap. For the edge state measurement, three sides of the lattice are wrapped with metal walls, and the remaining side is covered with absorbers. In this way, only clockwise (anticlockwise) propagation excited at point 1 (2) can be detected. The measured edge transmissions show a huge difference between forward and backward transport in the gap, indicating the existence of topological one-way edge states. We also map out the field distributions inside the waveguide, showing how the unidirectional edge state traveling through two 90° sharp corners without reflection (it is then absorbed after impinging on the microwave absorber). Due to the intrinsic absorptive loss of the gyromagnetic material, some dissipation of the edge state is observed during propagation.

Next, we fabricate an amorphous PTI with DI = 0.1 and characterize it using the same experimental setup. In the bulk transmission measurements, we observe a significant dip in both forward and backward transmission between 10.6 GHz to 11.4 GHz, indicating a mobility gap at frequencies close to the crystalline counterpart (Fig. 2**d**). In the edge measurements, we observe a huge difference between forward and backward transmission in the frequency range of the mobility gap (Fig. 2**d**). Mapping out the field distributions reveals a unidirectional edge state propagating clockwise. These experimental results are consistent with numerical calculations (see Fig. 1**d** and Supplementary Information), indicating that the localization length is extremely short from ~10.5 GHz to ~11.5 GHz. Since the amorphous PTI lacks periodicity, it lacks a properly defined momentum-space bandstructure; to characterize the topology, we adapt the Bott index (see Supplementary Information), which acts like the Chern number but can be applied in real space[27, 29]. As shown in Fig. 2**a**, the Bott index has a nontrivial value of 1 (equivalent to the Chern number for the earlier crystalline PTI) within the mobility gap. All of these results – the bulk gap, one-way chiral edge transport, first-principles calculations of the mobility gap, and the Bott index – are in excellent agreement, pointing to the existence of topologically protected edge states in the amorphous PTI.

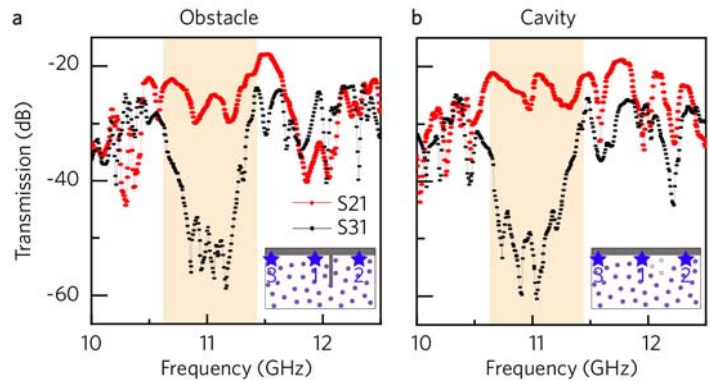

**Figure 3. Robust chiral edge propagation in amorphous PTI with defects.** (a) Measured edge transmission in the presence of a large obstacle. Inset: schematic of the experimental setup, where the source (1) and probe (2 or 3) antennas are marked as stars. Grey bars represent the metallic obstacle and boundary. The length of the obstacle is $3a$. (b) Measured edge transmission in the presence of a large cavity. Inset: schematic of the experimental setup. The light grey dots denote three removed gyromagnetic rods.

To verify the robustness of the edge states in the amorphous PTI, we introduce defects along the edges. Two types of defects were tested. In the first case, a rectangle aluminum obstacle is placed at the edge to block the edge propagation (Fig 3**a**). In the second case, three gyromagnetic rods are removed to create a large air cavity (Fig. 3**b**). We then measure the edge transmission. In both cases, we find large differences between forward and backward transmissions in the frequency range of the mobility gap, indicating that the defects do not cause backscattering.

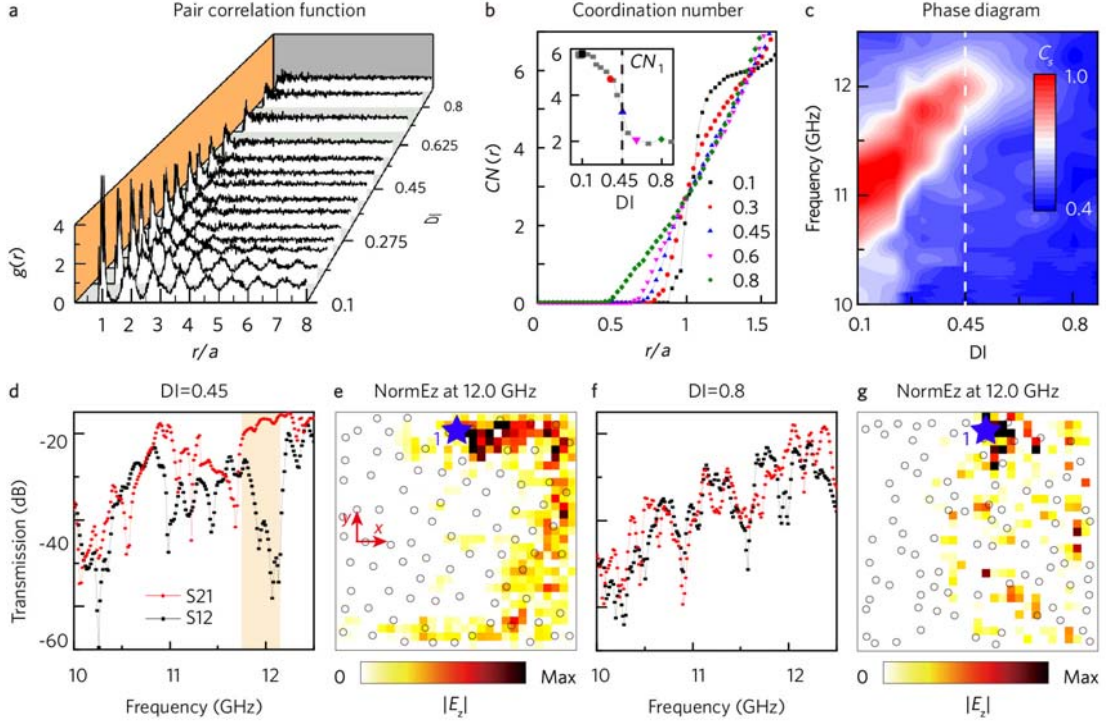

**Figure 4. Extinction of topological states in amorphous photonic lattices.** (a) Pair correlation function with different DI. (b) Running coordination number of photonic lattices with DI = 0.1- 0.8. Inset: coordination numbers of the first pair correlation function peak ($CN_1$). (c) Numerically-calculated empirical parameter $C_s$. Dotted lines (black in (b) and white in (c)) represent the critical DI approaching short-range order threshold. (d)-(g) Measured transmissions and |$E_z$| field distributions of the edge states in photonic lattice samples with DI = 0.45 and 0.8, respectively. Experimental setup is the same as that in Fig. 2. The orange region in (d) denotes the corresponding numerically-calculated topological region.

Next, we study the effects of the glass transition. It should be noted that the nature of the glass transition in real amorphous materials remains poorly understood, despite the extensive theoretical and experimental studies[31, 32]. Using the lattice generation procedure detailed above, Fig. 4**a** plots how the pair correlation function evolves with DI. To help locate the glass transition, we calculate the running coordination number[43]

$$CN(r) = \frac{2\pi N}{A} \int_0^r g(r) r dr \tag{2}$$

As shown in Fig. 4**b**, $CN(r)$ changes from a step-like curve to a smooth one with increasing DI. Integrating to the first minimum (a discontinuity) of $g(r)$ gives the coordination number of the nearest neighbors, denoted as $CN_1$ (inset of Fig. 4**b**), which represents the average local connectivity of each lattice site. As DI increases, $CN_1$ drops from a ~ 6 (similar to the crystalline case) to ~2 (similar to a liquid). Around the critical value of DI = 0.45, $CN_1$ drops very quickly, suggesting a glass transition[31, 44]. Thereafter, $CN_1$ converges to ~ 2, indicating the completion of the glass transition.

We used first-principle simulations to investigate the interplay between short-range order and topological protection. In the simulations, the photonic lattices are surrounded by PEC boundaries, and a point source is placed near the boundaries. Based on the numerical field distributions, we calculate an empirical parameter[26]

$$C_s = \frac{\int_{\Pi_s} \varepsilon(x,y)dxdy}{\int_{\Pi} \varepsilon(x,y)dxdy} \qquad (3)$$

where $\varepsilon$ is the electromagnetic (EM) energy density, $\Pi$ is the whole area of the photonic lattice, and $\Pi_s$ is the area one free-space wavelength away from the PEC boundary. When the system hosts topological edge states, they tend to be localized in $\Pi_s$, so $C_s$ is close to unity. The plot of $C_s$ versus DI is shown in Fig. 4**c**. For DI = 0.1, $C_s$ is close to unity within the frequency window corresponding to the mobility gap, consistent with the previous results. Upon increasing DI beyond the critical value of 0.45, i.e., around the glass transition, the short-range order quickly decreases, and the frequency window (the high $C_s$ region marked in red in Fig. 4**c**) shrinks rapidly to zero.

To verify these findings experimentally, we fabricate two samples with DI = 0.45 and DI = 0.8, and measure the edge transmission and the electric field distribution using the same setup as in Fig. 2. For DI = 0.45, the topological frequency window shrinks to a narrow range of 11.7 GHz -12.2 GHz (Fig. 4**c**), and the edge states are only weakly confined to the edge (Fig. 4**d**). For DI = 0.8, past the glass transition, there is no sign of the topological edge states in the transmission or field distribution measurements; the numerically-calculated localization length shows small fluctuations (Fig. 1**d**), suggesting the closing of the mobility gap.

We thus experimentally realized amorphous PTIs that lack long-range order but preserve short-range order. Using microwave measurements, we directly observed the bulk mobility gap and the unidirectional propagation of topological edge states, which is robust against defects and disorders. By gradually deforming the amorphous lattice into a liquid-like lattice through the glass transition, we observed the closing of the mobility gap and the disappearance of the topological edge states. These results illustrate the key role of short-range order in the formation of the topological edge states. These insights may be useful for realizing amorphous topological insulators in other physical settings such as acoustics. It would also be interesting to explore other types of non-crystalline photonic topological materials, such as topological random lasers.

**Methods**

**Sample and experimental measurement.** The yttrium iron garnet (YIG) ferrite cylinder rods have relative permittivity 13, dielectric loss tangent 0.0002, radius 2.2 mm, and height 4 mm. The saturation magnetization was measured to be $M_s$ = 1780 Gauss, and the gyromagnetic resonance loss width to be 35 Oe. In the microwave measurements, a static magnetic field generated by an electromagnet is applied perpendicular to the waveguide, producing a strong gyromagnetic response in the ferrite rods. The spatial non-uniformity of the magnetic field is less than 2% in the sample region.

**Simulation.** The band structure, bulk/edge transmissions, and field distributions are simulated using the finite element software COMSOL Multiphysics. For all the band structure calculations, the frequency dispersion and material losses are neglected for simplicity. The relative permeability tensor of the gyromagnetic materials is taken to be

$$\tilde{\mu} = \begin{bmatrix} 0.6734 & 0.6304i & 0 \\ -0.6304i & 0.6734 & 0 \\ 0 & 0 & 1 \end{bmatrix}$$

which is the experimentally-obtained value at 11 GHz with a 0.2 T static magnetic field along the -$z$ direction. The effects of dispersion are negligible since both the permittivity and the permeability vary only slightly in the considered frequency band.

**References**

1. Rechtsman, M.C., Zeuner, J.M., Plotnik, Y., Lumer, Y., Podolsky, D., Dreisow, F., Nolte, S., Segev, M. & Szameit, A. Photonic Floquet topological insulators. *Nature* **496**, 196-200 (2013).
2. Khanikaev, A.B., Mousavi, S.H., Tse, W.K., Kargarian, M., MacDonald, A.H. & Shvets, G. Photonic topological insulators. *Nature Materials* **12**, 233-239 (2013).
3. Lu, L., Joannopoulos, J.D. & Soljačić, M. Topological photonics. *Nature Photonics* **8**, 821-829 (2014).
4. Khanikaev, A.B. & Shvets, G. Two-dimensional topological photonics. *Nature Photonics* **11**, 763-773 (2017).
5. Ozawa, T., Price, H.M., Amo, A., Goldman, N., Hafezi, M., Lu, L., Rechtsman, M.C., Schuster, D., Simon, J., Zilberberg, O. & Carusotto, I. Topological photonics. *Reviews of Modern Physics* **91**, 015006 (2019).
6. Bandres, M.A., Wittek, S., Harari, G., Parto, M., Ren, J., Segev, M., Christodoulides, D.N. &

Khajavikhan, M. Topological insulator laser: Experiments. *Science* **359**, eaar4005 (2018).

7. Harari, G., Bandres, M.A., Lumer, Y., Rechtsman, M.C., Chong, Y.D., Khajavikhan, M., Christodoulides, D.N. & Segev, M. Topological insulator laser: Theory. *Science* **359**, eaar4003 (2018).
8. Bahari, B., Ndao, A., Vallini, F., El Amili, A., Fainman, Y. & Kante, B. Nonreciprocal lasing in topological cavities of arbitrary geometries. *Science* **358**, 636-640 (2017).
9. Hafezi, M., Demler, E.A., Lukin, M.D. & Taylor, J.M. Robust optical delay lines with topological protection. *Nature Physics* **7**, 907-912 (2011).
10. Haldane, F.D. & Raghu, S. Possible realization of directional optical waveguides in photonic crystals with broken time-reversal symmetry. *Physical Review Letters* **100**, 013904 (2008).
11. Raghu, S. & Haldane, F.D.M. Analogs of quantum-Hall-effect edge states in photonic crystals. *Physical Review A* **78**, 033834 (2008).
12. Wang, Z., Chong, Y.D., Joannopoulos, J.D. & Soljacic, M. Reflection-free one-way edge modes in a gyromagnetic photonic crystal. *Physical Review Letters* **100**, 013905 (2008).
13. Wang, Z., Chong, Y., Joannopoulos, J.D. & Soljacic, M. Observation of unidirectional backscattering-immune topological electromagnetic states. *Nature* **461**, 772-775 (2009).
14. Poo, Y., Wu, R.X., Lin, Z., Yang, Y. & Chan, C.T. Experimental realization of self-guiding unidirectional electromagnetic edge states. *Physical Review Letters* **106**, 093903 (2011).
15. Chen, W.J., Jiang, S.J., Chen, X.D., Zhu, B., Zhou, L., Dong, J.W. & Chan, C.T. Experimental realization of photonic topological insulator in a uniaxial metacrystal waveguide. *Nature Communications* **5**, 5782 (2014).
16. Cheng, X., Jouvaud, C., Ni, X., Mousavi, S.H., Genack, A.Z. & Khanikaev, A.B. Robust reconfigurable electromagnetic pathways within a photonic topological insulator. *Nature Materials* **15**, 542-548 (2016).
17. Wu, X., Meng, Y., Tian, J., Huang, Y., Xiang, H., Han, D. & Wen, W. Direct observation of valley-polarized topological edge states in designer surface plasmon crystals. *Nature Communications* **8**, 1304 (2017).
18. Yves, S., Fleury, R., Berthelot, T., Fink, M., Lemoult, F. & Lerosey, G. Crystalline metamaterials for topological properties at subwavelength scales. *Nature Communications* **8**, 16023 (2017).
19. Gao, F., Xue, H., Yang, Z., Lai, K., Yu, Y., Lin, X., Chong, Y., Shvets, G. & Zhang, B. Topologically protected refraction of robust kink states in valley photonic crystals. *Nature Physics* **14**, 140-144 (2018).
20. Noh, J., Huang, S., Chen, K.P. & Rechtsman, M.C. Observation of photonic topological valley Hall edge states. *Physical Review Letters* **120**, 063902 (2018).
21. Barik, S., Karasahin, A., Flower, C., Cai, T., Miyake, H., DeGottardi, W., Hafezi, M. & Waks, E. A topological quantum optics interface. *Science* **359**, 666-668 (2018).
22. Shalaev, M.I., Walasik, W., Tsukernik, A., Xu, Y. & Litchinitser, N.M. Robust topologically protected transport in photonic crystals at telecommunication wavelengths. *Nature Nanotechnology* **14**, 31-34 (2019).
23. Yang, Y., Gao, Z., Xue, H., Zhang, L., He, M., Yang, Z., Singh, R., Chong, Y., Zhang, B. & Chen, H. Realization of a three-dimensional photonic topological insulator. *Nature* **565**, 622-626 (2019).
24. He, X.T., Liang, E.T., Yuan, J.J., Qiu, H.Y., Chen, X.D., Zhao, F.L. & Dong, J.W. A silicon-on-insulator slab for topological valley transport. *Nature Communications* **10**, 872

(2019).

25. Keller, J. & Ziman, J. Long range order, short range order and energy gaps. *Journal of Non-Crystalline Solids* **8**, 111-121 (1972).
26. Liu, C., Gao, W., Yang, B. & Zhang, S. Disorder-induced topological state transition in photonic metamaterials. *Physical Review Letters* **119**, 183901 (2017).
27. Titum, P., Lindner, N.H., Rechtsman, M.C. & Refael, G. Disorder-induced Floquet topological insulators. *Physical Review Letters* **114**, 056801 (2015).
28. Stutzer, S., Plotnik, Y., Lumer, Y., Titum, P., Lindner, N.H., Segev, M., Rechtsman, M.C. & Szameit, A. Photonic topological Anderson insulators. *Nature* **560**, 461-465 (2018).
29. Bourne, C. & Prodan, E. Non-commutative Chern numbers for generic aperiodic discrete systems. *Journal of Physics A: Mathematical and Theoretical* **51**, 235202 (2018).
30. Stachurski, Z.H. On structure and properties of amorphous materials. *Materials* **4**, 1564-1598 (2011).
31. Berthier, L. & Biroli, G. Theoretical perspective on the glass transition and amorphous materials. *Reviews of Modern Physics* **83**, 587 (2011).
32. Amann-Winkel, K., Böhmer, R., Fujara, F., Gainaru, C., Geil, B. & Loerting, T. Colloquium: Water's controversial glass transitions. *Reviews of Modern Physics* **88**, 011002 (2016).
33. Yang, B., Zhang, H., Wu, T., Dong, R., Yan, X. & Zhang, X. Topological states in amorphous magnetic photonic lattices. *Physical Review B* **99**, 045307 (2019).
34. Mansha, S. & Chong, Y.D. Robust edge states in amorphous gyromagnetic photonic lattices. *Physical Review B* **96**, 121405 (2017).
35. Agarwala, A. in *Excursions in Ill-Condensed Quantum Matter* 61-79 (Springer, 2019).
36. Agarwala, A., Juricic, V. & Roy, B. Higher Order Topological Insulators in Amorphous Solids. *arXiv:1902.00507* https://arxiv.org/abs/1902.00507 (2019).
37. Mitchell, N.P., Nash, L.M., Hexner, D., Turner, A.M. & Irvine, W.T. Amorphous topological insulators constructed from random point sets. *Nature Physics* **14**, 380 (2018).
38. Gao, G.-J., Bławzdziewicz, J. & O'Hern, C.S. Frequency distribution of mechanically stable disk packings. *Physical Review E* **74**, 061304 (2006).
39. Torquato, S. Perspective: Basic understanding of condensed phases of matter via packing models. *The Journal of Chemical Physics* **149**, 020901 (2018).
40. Xia, C., Li, J., Cao, Y., Kou, B., Xiao, X., Fezzaa, K., Xiao, T. & Wang, Y. The structural origin of the hard-sphere glass transition in granular packing. *Nature communications* **6**, 8409 (2015).
41. Luo, W., Sheng, H. & Ma, E. Pair correlation functions and structural building schemes in amorphous alloys. *Applied Physics Letters* **89**, 131927 (2006).
42. Ding, J., Ma, E., Asta, M. & Ritchie, R.O. Second-nearest-neighbor correlations from connection of atomic packing motifs in metallic glasses and liquids. *Scientific Reports* **5**, 17429 (2015).
43. Palombo, M., Gabrielli, A., Servedio, V., Ruocco, G. & Capuani, S. Structural disorder and anomalous diffusion in random packing of spheres. *Scientific Reports* **3**, 2631 (2013).
44. Lewis, L.J. Atomic dynamics through the glass transition. *Physical Review B* **44**, 4245 (1991).